\documentclass[twocolumn]{aastex7}
\usepackage{amsmath}
\usepackage{bbm,bm}
\usepackage{multirow}
\usepackage{colortbl}
\usepackage{outlines}
\usepackage{comment,color}
\usepackage{CJKutf8}

\usepackage{xcolor}

\graphicspath{{./}{figures/}}

\shorttitle{HLX--1 as a Repeating Partial TDE}
\shortauthors{Yu et al.}

\begin{document}

\title{HLX--1 as a Repeating Partial Tidal Disruption Event around an Intermediate-Mass Black Hole}

\author[0009-0004-6973-3955]{Fangyuan Yu \begin{CJK*}{UTF8}{gbsn}(俞方远)\end{CJK*}}
\affiliation{Department of Astrophysical Sciences, Princeton University, 4 Ivy Lane, Princeton, NJ 08544, USA}
\email[show]{fy6204@princeton.edu}

\author{Andrew Mummery}
\affiliation{School of Natural Sciences, Institute for Advanced Study, 1 Einstein Drive, Princeton, NJ 08540, USA}
\email{amummery@ias.edu}

\author[0000-0002-5063-0751]{Muryel Guolo}
\affiliation{School of Natural Sciences, Institute for Advanced Study, 1 Einstein Drive, Princeton, NJ 08540, USA}
\affiliation{Bloomberg Center for Physics and Astronomy, Johns Hopkins University, 3400 N. Charles St., Baltimore, MD 21218, USA}
\email{mguolop1@jhu.edu}

\begin{abstract}
ESO 243$-$49 HLX--1 is one of the first known compelling intermediate-mass black hole (IMBH) candidates, but the origin of its recurrent X-ray outbursts remains unsettled. In this work we fit the HLX--1 outbursts simultaneously with a relativistic thin-disk time-dependent model. The black-hole and disk-geometry parameters are tied across epochs, while the injected mass and timing parameters are fitted separately for each outburst. We also include a contemporaneous \textit{HST} UV/optical/IR spectral energy distribution obtained near the peak of one outburst, which constrains the global temperature and radius scales of the disk. The soft-state data are consistent with a common global solution, with black hole mass $\log_{10}(M_{\bullet}/M_\odot)=4.5\pm0.2$ and disk formation radius $\log_{10}(r_0/r_g)=3.8\pm0.3$. The differences between outbursts are mainly described by modest changes in disk mass and viscous timescales. The inferred injected masses are typically $\sim 10^{-3}\,M_\odot$ per outburst, requiring a total fuel supply of at least a few $\times 10^{-3}\,M_\odot$ across the observed sequence. This mass scale, and the disk size, are consistent with repeated partial stripping of a low-mass donor. A simple Keplerian mapping of the flare gaps and fitted disk radius implies a highly eccentric orbit with a pericenter just outside the donor's tidal radius. These results support a repeating partial tidal disruption interpretation for the origin of HLX--1.
\end{abstract}


\section{Introduction}
\label{sec:intro}

Intermediate-mass black holes (IMBHs) occupy the mass range between stellar-mass and supermassive black holes, but their existence remains much less securely established than that of either endpoint population \citep[e.g.,][]{Mezcua2017IJMPD,Greene2020ARAA}. Dynamical searches are difficult beyond the Local Group, and accretion signatures can overlap with those of unusual stellar-mass systems or weakly accreting galactic nuclei. Tidal disruption events (TDEs) provide another route: they can temporarily illuminate otherwise quiescent black holes, and the flare properties scale with the black hole mass and disk properties \citep[e.g.,][for a TDE review]{Gezari2021ARAA}. Recent population studies and disk-based modeling support an evolving accretion-flow interpretation for many UV/optical and X-ray TDEs \citep{vanVelzen19,MummeryBalbus2020, Mummery2024MNRAS,Mummery2025MNRAS,Guolo2026b}. Off-nuclear TDEs are particularly useful for this purpose because they can reveal accreting black holes outside of the primary galactic nucleus, including wandering black holes, stripped satellite nuclei, or massive star clusters \citep{Lin2018NatAs,Yao2025ApJ,Jin2025,Stein2026,Guolo2026}.

ESO~243$-$49 HLX--1 connects these issues directly. It was discovered as a hyperluminous X-ray source offset from the nucleus of ESO~243$-$49 and became one of the strongest IMBH candidates known \citep[e.g.,][]{Farrell2009Nature}. Subsequent \textit{Chandra} and \textit{Swift} follow-up improved the source localization and established its multiwavelength context \citep{Webb2010ApJL}. Its peak X-ray luminosity reaches $\sim 10^{42}\ {\rm erg\ s^{-1}}$, above the luminosities of ordinary Galactic X-ray binaries, while its soft, disk-dominated states, state-transition behavior, and evidence for jet activity resemble canonical black-hole accretion phenomenology \citep{Godet2009ApJ,Davis2011ApJ,Servillat2011ApJ,Webb2012Science}. Optical monitoring has shown that part of the optical emission varies with the accretion state, supporting a disk contribution to the counterpart light \citep{Webb2014ApJL}. The IMBH interpretation is now widely favored, but two problems have remained debated in the literature.

The first is what powers the repeating outbursts of HLX--1. The second is the nature of the UV/optical emission, in particular whether a stellar cluster is present and, if so, what its properties are. This paper addresses the first; a follow-up paper will address the second.

The long-term monitoring history of HLX--1 provides unusually strong constraints on the outburst mechanism. \citet{Yan2015ApJ} showed that the source underwent a sequence of bright X-ray outbursts whose recurrence time increased over the observed baseline, while the outburst durations decreased and the peak luminosities remained comparatively stable. A successful model must therefore explain not only repeated triggering, but also the linked evolution of recurrence time, duration, and energetics across multiple cycles.

Several mechanisms have been proposed. Thermal-viscous disk-instability models can reproduce some aspects of the phenomenology, but straightforward applications struggle to match the observed recurrence times, luminosities, and outburst durations at the distance of ESO~243$-$49 \citep{Lasota2011ApJ}. Feedback-regulated accretion cycles, including radiation-pressure-driven oscillations and wind- or irradiation-mediated limit cycles, have also been explored \citep{Wu2016ApJ,Soria2017MNRAS}. Other interpretations invoke episodic mass supply from an eccentric donor orbit or superorbital/geometric modulation in a super-Eddington system \citep{Lasota2011ApJ,King2014MNRAS,vanderHelm2016MNRAS}. These models explain different parts of the data, but none provides an unambiguous account of the full multi-epoch behavior.

At the same time, the astrophysical case for repeating partial tidal disruption has strengthened substantially. Both observations and theory now support the idea that a star on a bound orbit can survive multiple close passages while losing mass repeatedly and producing a sequence of flares \citep[e.g.,][]{Lin2024AT2022dbl,Bandopadhyay2024ApJ,Somalwar2025AT2020vdq}. Hydrodynamic calculations show that partial disruptions can alter the stellar structure, remnant orbit, and stripped mass \citep[e.g.,][]{Ryu2020_pTDE,Cufari2023MNRAS,Chen2024ApJ,Sharma2024MNRAS,Liu2025ApJ}, while theoretical work has emphasized that modest changes in orbital energy at high eccentricity can produce substantial period evolution even when the pericenter geometry remains comparatively stable \citep{Linial2024MNRAS,Bandopadhyay2024ApJ}.
These developments motivate testing the repeating partial-TDE scenario against the multi-epoch data of HLX--1.

Such a test is particularly constraining because a repeated-injection model imposes cross-epoch consistency requirements. Global quantities such as the black-hole mass, viewing geometry, and characteristic disk-formation scale should remain approximately fixed, whereas the injected mass reservoir and ``viscous'' timescale may vary between outbursts. In addition, one available UV/optical epoch is contemporaneous with the soft X-ray state, providing a broadband constraint on the temperature and radial scales of the flow that is unavailable from the X-ray light curves alone. This motivates a simultaneous multi-epoch analysis in which shared and epoch-dependent parameters are inferred separately.

In this paper, we model the HLX--1 outbursts with a relativistic thin-disk time-dependent framework and fit all epochs simultaneously. The black-hole and disk-geometry parameters are shared across epochs, while the injected mass and timing parameters vary between outbursts. We use this framework to
separate the accretion properties shared across outbursts from the epoch-dependent fuel and timing parameters, and to connect the inferred disk scales to
a repeating partial-TDE interpretation.

The paper is organized as follows. In Section~\ref{sec:methods} we describe the data sets, disk model, and fitting method. In Section~\ref{sec:results} we present the fits, parameter constraints, and derived quantities. In Section~\ref{sec:discussion} we discuss the physical interpretation, relation to previous models, and limitations.

\section{Methods}
\label{sec:methods}

\subsection{Data Reduction}
\label{sec:data_reduction}

In this paper, we use X-ray data from the X-ray Telescope (XRT) onboard the {\it Neil Gehrels Swift Observatory} \citep{Burrows2005SSRv}, as well as UV, optical, and IR photometry from the Wide Field Camera 3 (WFC3) onboard the {\it Hubble Space Telescope} \citep{Kimble2008}.

The {\it Swift}/XRT count-rate light curve was produced using the Swift UK online tools \citep{Evans2009}, with one bin per observation. We converted the count rate to unabsorbed 0.3--10~keV flux using the high/soft-state spectral model of \citet{GuoloMummery2025}, since we focus here on the soft state.

These were corrected for absorption using a Galactic hydrogen-equivalent column density of $N_{\rm H, gal} = 2\times10^{20} \ {\rm cm^{-2}}$ \citep{HI4PI2016} and their best-fitting intrinsic column density of $N_{\rm H, int} = 8\times10^{20} \ {\rm cm^{-2}}$. The fluxes were then converted to luminosities assuming a redshift of $z = 0.022$.

The observed UV/optical/IR data from filters F140LP, F300X, F390W, F555W, F775W, and F160W were taken from (\citealt{Soria2017MNRAS}, their Table~1). These were corrected for Galactic extinction using $E(B-V)_{\rm gal} = 0.02 $ from \citet{SchlaflyFinkbeiner2011} and the \citet{Cardelli1989} extinction law. Intrinsic dust extinction was corrected using $E(B-V)_{\rm int} = 0.09$, the best-fitting value from \citet{GuoloMummery2025}, which assumes a Galactic-like gas-to-dust ratio \citep{Ozel2012}, and the \citet{Calzetti2000} attenuation curve. We use only the brightest UV/optical/IR data taken in September 2010, which are known to be unambiguously dominated by disk emission \citep{Soria2017MNRAS,GuoloMummery2025}.

\subsection{Fitting Method}
\label{sec:fitting}

\subsubsection{Relativistic viscous evolution model}
\label{sec:disk_model}

We model the time-dependent evolution of the accretion flow using the analytical framework implemented in the \texttt{FitTeD} package \citep{fitted}\footnote{The public \texttt{FitTeD} repository is available at \url{https://github.com/fittingtransientswithdiscs/FitTeD}}. We consider a geometrically thin, optically thick disk evolving in the equatorial plane of a Kerr black hole of mass $M_{\bullet}$ and dimensionless spin $a_\bullet$. We adopt standard near-equatorial Boyer--Lindquist coordinates $(r,\phi,z,t)$, and denote the fluid four-velocity on circular orbits by $U^\mu$ \citep[with covariant counterpart $U_\mu$, e.g.,][]{Balbus2017,MummeryBalbus2020}.

Angular momentum transport is described by an anomalous (turbulent) stress tensor, whose dominant component in a thin disk is $W^{r}{}_{\phi}$. Following \citet{MummeryBalbus2020}, it is convenient to introduce
\begin{equation}
\zeta \equiv \frac{\sqrt{g}\,\Sigma\,W^{r}{}_{\phi}}{U^{0}},
\label{eq:zeta_def}
\end{equation}
where $\Sigma(r,t)$ is the height-integrated surface density, $g$ is the determinant of the mid-plane Kerr metric, and for our coordinate choice $\sqrt{g}=r$.

The general relativistic thin-disk evolution equation may be written directly for $\Sigma$ as \citep{Balbus2017}
\begin{equation}
\frac{\partial \Sigma}{\partial t}=\frac{1}{rU^{0}}\frac{\partial}{\partial r}
\left[
\frac{U^{0}}{U'_{\phi}}
\frac{\partial}{\partial r}
\left(
\frac{r\Sigma W^{r}{}_{\phi}}{U^{0}}
\right)
\right],
\label{eq:diffusion_pde}
\end{equation}
where a prime denotes a radial derivative. Equivalently, in terms of $\zeta$ the governing equation can be cast in a diffusion-like form:
\begin{equation}
\frac{\partial \zeta}{\partial t}=W\,\frac{\partial}{\partial r}\left(\frac{U^{0}}{U'_{\phi}}\frac{\partial \zeta}{\partial r}\right),
\end{equation}
with
\begin{equation}
W \equiv \frac{1}{(U^{0})^{2}}\left(W^{r}{}_{\phi}+\Sigma\,\frac{\partial W^{r}{}_{\phi}}{\partial \Sigma}\right).
\end{equation}
For convenience, we impose a vanishing-stress inner boundary condition at the innermost stable circular orbit (ISCO) radius $r_{\rm ISCO}$, $W^{r}{}_{\phi}(r_{\rm ISCO},t)=0$, which is equivalently $\zeta(r_{\rm ISCO},t)=0$.

In \texttt{FitTeD}, the disk is initialized as a narrow ring with total mass $M_{\rm d}$ at a characteristic radius $r_0>r_{\rm ISCO}$. \texttt{FitTeD} uses this solution to compute the evolving surface-density profile, from which the temperature profile and emergent spectrum are obtained (see \citealt{fitted} for the full derivation and implementation details). Schematically, the solution may be written as
\begin{equation}
\begin{aligned}
\Sigma(\tau, x)\ \propto\ \tau^{-1}\,
&\exp\!\left[-\frac{f_\alpha(x)^2+f_\alpha(x_0)^2}{4\tau}\right]\, \\
&I_{1/(4\alpha)}\!\left(\frac{f_\alpha(x)\,f_\alpha(x_0)}{2\tau}\right),
\end{aligned}
\label{eq:greens_function}
\end{equation}
where $I_\nu$ is the modified Bessel function, $x \equiv 2 r/r_{\rm ISCO}$ is a dimensionless radial coordinate (with $x_0$ corresponding to $r=r_0$), $\tau$ is the dimensionless evolution variable related to the shifted time $(t+t_0)/t_{\rm visc}$, and $f_\alpha(x)$ is a function of the dimensionless radius (see \citealt{fitted} for more details). For the default stress prescription in \texttt{FitTeD}, the Bessel-function order is $1/(4\alpha)=1/3$.

Given the evolving stress (or equivalently $\zeta$), we compute the local disk surface temperature by balancing local shear dissipation with blackbody cooling:
\begin{equation}
\sigma_{\rm SB} T_{\rm eff}^4(r,t)=-\frac{U^{0}U_{\phi}}{2r}\,(\ln\Omega)'\,\zeta(r,t),
\label{eq:Teff_from_zeta}
\end{equation}
where $\Omega \equiv {\rm d}\phi/{\rm d}t = U^{\phi}/U^{0}$.

Finally, we compute the observed emission by integrating the local color-corrected blackbody emission over the disk surface, adopting the temperature-dependent $f_{\rm col}(T)$ prescription of \citet{Done2012MNRAS}.
In the present implementation we adopt the Newtonian limit, so that
\begin{equation}
L_\nu(t)
=
4\pi \cos\iota
\int_{r_{\rm ISCO}}^{r_{\rm out}}
B_\nu\!\left(f_{\rm col} T_{\rm eff}\right)\,
f_{\rm col}^{-4}\,
2\pi r\,{\rm d}r.
\label{eq:obs_flux}
\end{equation}
The disk evolution retains the relativistic Kerr orbital dynamics, while the emergent spectrum is evaluated with the Newtonian surface integral of Equation~\ref{eq:obs_flux}. For HLX--1, \citet{GuoloMummery2025} showed that including full relativistic transfer changes the inferred posterior constraints by less than $1\sigma$ (their Figure~12).

\subsubsection{Data selection and all-epoch inference}
\label{sec:data_selection}
\label{sec:all_epoch_fit}

We divide the \textit{Swift}/XRT monitoring of HLX--1 into discrete epochs, each corresponding to one X-ray outburst separated from the next by a low-state interval identified from the hardness-ratio evolution. Several \textit{HST} UV/optical observations exist, but only the September 2010 data are contemporaneous with the soft X-ray state. We therefore fit these data jointly with the Epoch~3 X-ray light curve.

For each epoch $i$, we use the fine-resolution $0.3$--$10$~keV light curve to define three characteristic times: the last secure low-state point, $t_{{\rm low},i}$; the first clear rising-outburst point, $t_{{\rm on},i}$; and the onset of the next hard state transition, $t_{{\rm tr},i}$, which marks the end of the fitting window. Only data within
\[
t_{{\rm on},i} < t < t_{{\rm tr},i}
\]
enter the X-ray likelihood (the fitted window for each epoch is shown as gray shaded bands in Figure~\ref{fig:xray_epochs}). The preceding window,
\[
t_{{\rm low},i} < t < t_{{\rm on},i},
\]
is used only to define the epoch-specific prior on the disk formation-time $t_{0,i}$. The hardness-ratio evolution is used as a consistency check for this selection, not as a fitted observable. We take $t_{{\rm tr},i}$ to mark the first sustained hardening away from the soft state, identified for each epoch from the hardness-ratio evolution rather than by a fixed numerical threshold.

We therefore restrict the fit to the soft-state portion of each outburst. The thin-disk evolution framework assumes a radiatively efficient, geometrically thin disk. After the source hardens and enters a lower-accretion-rate state, the thin-disk light-curve model is no longer expected to apply \citep[e.g.,][]{Narayan&Yi1995ApJ,Done2007AARv,YuanNarayan2014}. We therefore exclude the late-time post-break measurements from the inference and show them only for comparison.

The X-ray measurements included in the fit are rebinned in two steps. First, we merge adjacent measurements toward a target count-rate signal-to-noise ratio of 2.5, with a nominal maximum bin width of 2~days and not merging across gaps longer than 1~day. We then merge additional neighboring bins until each epoch contains at most 16 bins, preventing densely sampled epochs from dominating by point count. Merged times, count rates, and luminosities are computed as inverse-variance-weighted means. The fine-resolution light curves are retained only for visualization.

We fit all selected epochs simultaneously. The shared parameters are
\begin{equation}
\boldsymbol{\theta}_{\rm g}
=
\left(
\log_{10}\frac{M_{\bullet}}{M_\odot},\,
a_\bullet,\,
\cos\iota,\,
\log_{10}\frac{r_0}{r_g}
\right),
\end{equation}
where $M_{\bullet}$ is the black-hole mass, $a_\bullet$ is the dimensionless spin, $\iota$ is the disk-observer inclination, and $r_0$ is the characteristic disk-formation radius. We sample the viewing geometry uniformly in $\cos\iota$ (isotropic prior) and require $r_0>r_{\rm ISCO}(a_\bullet)$.
In our fiducial fit $r_0$ is shared across epochs. For the parameters inferred here the fitted $0.3$--$10$~keV band is emitted from the innermost tens of gravitational radii, whereas the contemporaneous \textit{HST} filters probe radii comparable to $r_0$, so X-ray-only epochs carry almost no information about the angular-momentum scale. We test this directly by repeating the all-epoch inference with $r_{0}$ free for every outburst, adopting the same broad top-hat prior $\log_{10}(r_0/r_g)\in[2.0,5.5]$ independently for each epoch while $M_{\bullet}$, $a_\bullet$ and the inclination remain shared. For comparison we also show earlier per-epoch fits in which one-dimensional kernel density estimates of the Epoch~3 marginals were adopted as priors on the global parameters. Both sets of results are discussed in Section~\ref{sec:disk_ptde}.
The epoch-dependent parameters are
\begin{equation}
\boldsymbol{\theta}_{i}
=
\left(
\log_{10}\frac{M_{{\rm d},i}}{M_\odot},\,
t_{{\rm visc},i},\,
t_{0,i}
\right),
\end{equation}
where $M_{{\rm d},i}$ is the injected disk mass, $t_{{\rm visc},i}$ is the viscous timescale, and $t_{0,i}$ is the disk formation-time offset for epoch $i$.

For each epoch, the model X-ray luminosity is obtained by integrating the evolving spectral luminosity over the observed bandpass,
\begin{equation}
\mu_{X,i}(t)
\equiv
L_{X,i}(t)
=
\int_{\nu_1}^{\nu_2} L_\nu(t)\,{\rm d}\nu,
\end{equation}
where $h\nu_1=0.3~{\rm keV}$ and $h\nu_2=10~{\rm keV}$. For the Epoch~3 UV/optical SED, we evaluate
\begin{equation}
\mu_{{\rm UV},3}(t_{\rm UV},\nu)
\equiv
\nu L_\nu(t_{\rm UV},\nu),
\end{equation}
at the observed UV/optical frequencies.

In the likelihood evaluation, $\mu_{X,i}(t)$ is computed from the evolution of a single ring injected in epoch~$i$ alone. Residual emission from earlier disks is included in the posterior-predictive visualization through disk stacking, in which the model light curve at time~$t$ during epoch~$i$ is constructed from the joint thermal profile of all disks injected up to and including that epoch. Specifically, each contributing disk $j\leq i$ is evaluated at its own age $t - t_{0,j}^{\rm abs}$ (where $t_{0,j}^{\rm abs}$ is the absolute formation time of epoch~$j$) to obtain a radial effective-temperature profile $T_j(r)$. These profiles are combined via
\begin{equation}
T_{{\rm joint}, i}^4(r) = \sum_{j\leq i} T_j^4(r),
\label{eq:T4stack}
\end{equation}
and the resulting joint temperature profile is passed through a single color-corrected SED computation (Equation~\ref{eq:obs_flux}) and integrated over the 0.3--10~keV band. The $T^4$ summation follows because the governing evolution equation is linear in $\zeta$ (so that sums of solutions remain solutions), and $T^4\propto \zeta$. The implications of this approximation are discussed in Section~\ref{sec:disk_caveats}.

Let the rebinned X-ray data for epoch $i$ be
\[
D_{X,i}
=
\{t_{i,n},\,y_{i,n},\,\sigma_{i,n}\}_{n=1}^{N_{X,i}},
\]
where $y_{i,n}$ is the observed $0.3$--$10$~keV luminosity and $\sigma_{i,n}$ its $1\sigma$ uncertainty. We adopt a Gaussian log-likelihood
\begin{equation}
\ln \mathcal{L}_{X,i}=-\frac{1}{2}\sum_{n=1}^{N_{X,i}}\Delta_{X,i,n}^2,
\end{equation}
with
\begin{equation}
\Delta_{X,i,n}\equiv\frac{y_{i,n}-\mu_{X,i}(t_{i,n}\mid \boldsymbol{\theta}_{\rm g},\boldsymbol{\theta}_i)}{\sigma_{i,n}}.
\end{equation}
For the Epoch~3 UV/optical measurements,
\[
D_{{\rm UV},3}=\{\nu_k,\,y_{{\rm UV},k},\,\sigma_{{\rm UV},k}\}_{k=1}^{N_{{\rm UV},3}},
\]
we similarly use
\begin{equation}
\begin{aligned}
&\ln \mathcal{L}_{{\rm UV},3}=-\frac{1}{2}\sum_{k=1}^{N_{{\rm UV},3}}\Delta_{{\rm UV},k}^2,\\
&\Delta_{{\rm UV},k}\equiv\frac{y_{{\rm UV},k}-\mu_{{\rm UV},3}(t_{\rm UV},\nu_k\mid \boldsymbol{\theta}_{\rm g},\boldsymbol{\theta}_3)}{\sigma_{{\rm UV},k}}.
\end{aligned}
\end{equation}

The full fit uses an epoch-balanced composite likelihood, defined by
\begin{equation}
\ln \mathcal{L}_{\rm tot}=\sum_i \frac{\ln \mathcal{L}_{X,i}}{N_{X,i}}+\frac{\ln \mathcal{L}_{{\rm UV},3}}{N_{{\rm UV},3}}.
\label{eq:composite_like}
\end{equation}
This weighting prevents the most densely sampled X-ray epochs from dominating purely by point count, while still allowing the Epoch~3 UV/optical SED to constrain the global temperature and radius scales.

The corresponding composite-likelihood posterior is
\begin{equation}
p\!\left(\boldsymbol{\theta}_{\rm g},\{\boldsymbol{\theta}_i\}\mid D\right)\propto p(\boldsymbol{\theta}_{\rm g})\prod_i p(\boldsymbol{\theta}_i)\,\mathcal{L}_{\rm tot}(D\mid \boldsymbol{\theta}_{\rm g},\{\boldsymbol{\theta}_i\}),
\end{equation}
where $D$ denotes the full multi-epoch X-ray data set together with the Epoch~3 UV/optical measurements.

We adopt broad top-hat priors: $\log_{10}(M_{\bullet}/M_\odot)\in[2.5,6]$, $\log_{10}(r_0/r_g)\in[2.0,5.5]$, $\cos\iota\in[0,1]$, and $a_\bullet\in(-0.999,0.999)$ for the global parameters, and $\log_{10}(M_{{\rm d},i}/M_\odot)\in[-5.5,-1.0]$ and $t_{{\rm visc},i}\in[0,120]\,{\rm d}$ for each epoch. The prior on $t_{0,i}$ is uniform over the onset window defined above, allowing the data within the fitted interval to determine the preferred disk age without forcing the model to match the first detected rise in detail. Unless noted otherwise, the posterior constraints do not accumulate at these hard bounds.

We sample the posterior with the affine-invariant ensemble MCMC sampler \texttt{emcee} \citep{ForemanMackey+2013}. We report posterior medians and central credible intervals, and visualize the marginalized projections with standard corner plots.

\section{Results}
\label{sec:results}

We present the results of the single direct fit to all selected X-ray epochs, with the UV/optical SED of Epoch~3 included simultaneously in the likelihood. In this framework, the black hole and disk-geometry parameters are shared across outbursts, while the inferred injected mass per outburst and timing parameters are allowed to vary from epoch to epoch. Epoch~3 is special in the observational sense that it provides the only contemporaneous UV/optical constraint.

The UV/optical data provide leverage on the global temperature and radius scales of the disk spectrum: in a thin-disk model the UV (outer) and X-ray (inner) bands are dominated by different radii of the evolving multi-temperature emission, so a joint constraint helps break key degeneracies (e.g. between $M_{\bullet}$ and $r_0$).

\subsection{Posterior-predictive fits}
\label{sec:ppd_fits}

\begin{figure*}[t]
\centering
\includegraphics[width=\textwidth]{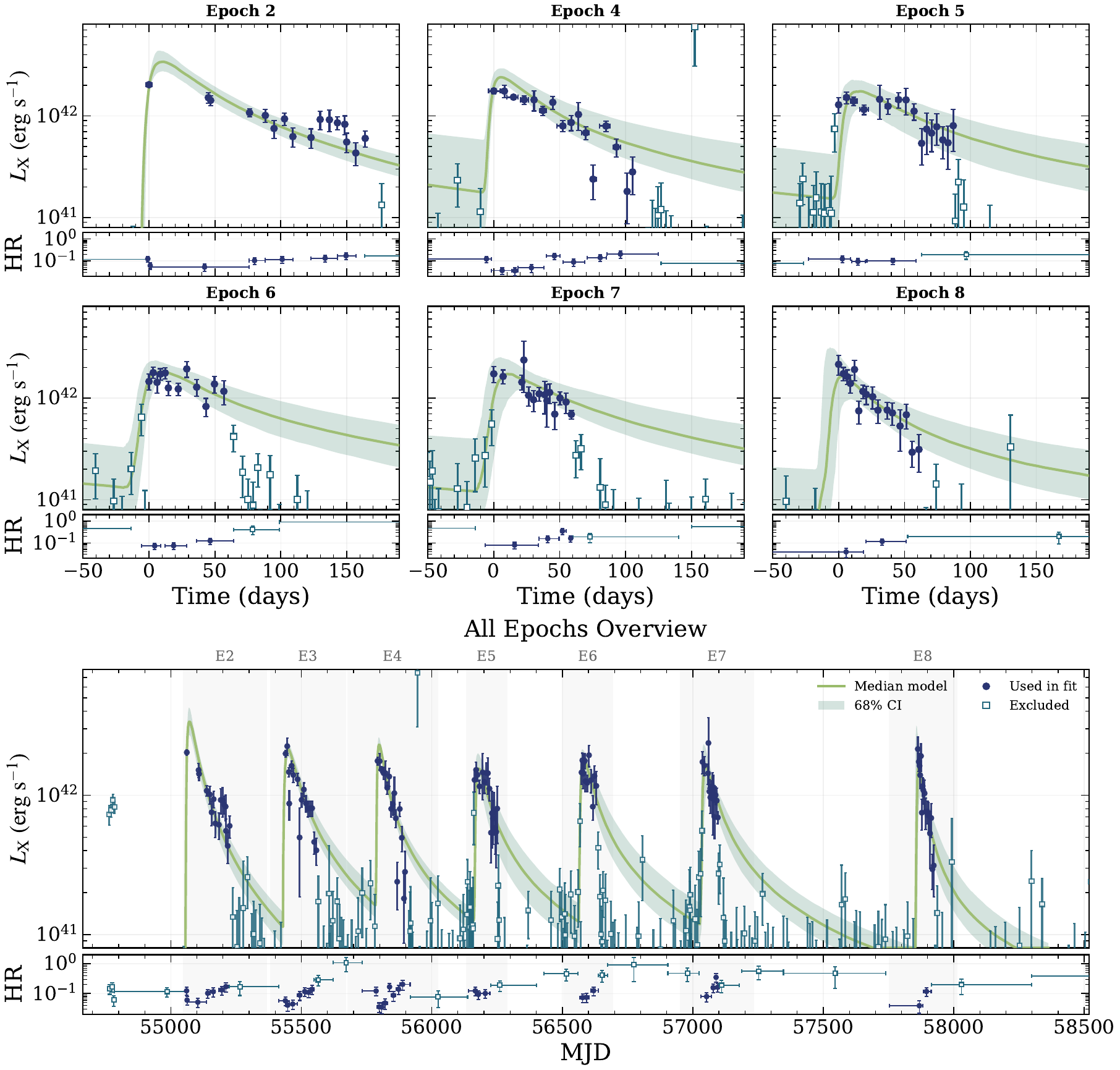}
\caption{
Multi-epoch X-ray evolution of HLX--1 and the corresponding posterior predictive model constraints for the adopted data selection. The top and middle rows show the individual outburst episodes (Epochs~2 and 4--8), plotted as a function of time in days relative to the reference time of each epoch. For each epoch, the upper panel shows the X-ray luminosity light curve, while the lower narrow panel shows the hardness ratio (HR) on a logarithmic scale. The bottom row presents the full long-term X-ray light curve as a function of MJD, together with the corresponding HR panel below. Gray shaded vertical bands mark the time span of each epoch. In all panels, filled circles denote data points included in the fit, while open squares indicate points excluded from the fit on the basis of spectral hardening (see hardness-ratio panels), marking the onset of the state transition. In the luminosity panels, the solid curve and shaded band show the median model prediction and the 68\% credible interval derived from the posterior samples, respectively. The HR measurements are shown for comparison only.
}
\label{fig:xray_epochs}
\end{figure*}

\begin{figure*}[t]
\centering
\includegraphics[width=\textwidth]{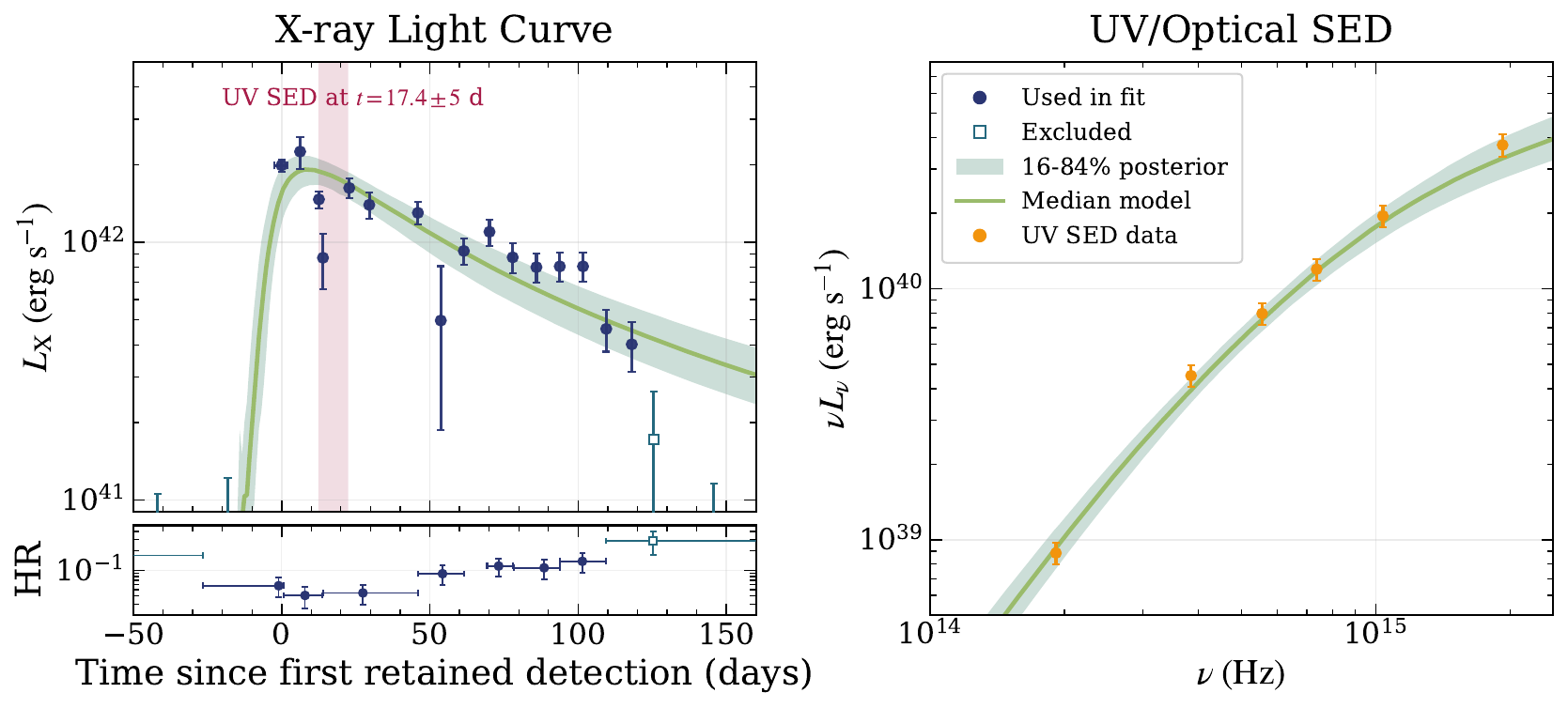}
\caption{
Epoch~3 posterior-predictive fit to HLX--1 in X-rays (left) and the UV/optical SED (right) within the direct all-epoch inference. In the left panel, filled circles denote rebinned X-ray data included in the fit and open squares denote excluded fine-resolution X-ray points; the solid curve shows the posterior median model, and the shaded band encloses the central 68\% credible interval from posterior predictive realizations. The vertical shaded region marks the time window of the UV observations used to construct the SED relative to the first X-ray detection included in the fit. In the right panel, points show the UV/optical SED with 1$\sigma$ errors, with the posterior median model and its 68\% credible interval shown by the solid curve and shaded band, respectively.
}
\label{fig:E3_uvx}
\end{figure*}

Figure~\ref{fig:xray_epochs} presents the posterior-predictive fits to the X-ray light curves for each epoch (except for Epoch~3, which is shown in Figure~\ref{fig:E3_uvx}), together with an overview in absolute time. These curves include the disk-stacking procedure described in Section~\ref{sec:fitting} (Equation~\ref{eq:T4stack}): at each time, the residual thermal contributions from all earlier disks are summed with the current epoch's emission. The cumulative residual from earlier disks contributes $\sim 15$--$20\%$ of each epoch's peak X-ray luminosity (posterior median), so the current outburst dominates while earlier disks provide a non-negligible underlying floor. Across all epochs, the model reproduces the characteristic fast-rise/slow-decay morphology over the time interval included in the fit. Where the decay is densely sampled (e.g. Epoch~2), the posterior-predictive band is narrow and closely tracks the curvature of the decline; where the rise and pre-peak coverage are sparse (e.g. Epoch~8), the predictive envelope broadens at early times, reflecting covariance with the onset parameter $t_0$ and the lack of strong constraints before the first detection.

In multiple epochs the light curve exhibits a terminal deviation from the smooth fitted soft-state evolution: a late-time point (or short sequence) drops below the extrapolation of the model. These measurements are accompanied by spectral hardening in the HR evolution and are therefore identified observationally with a transition out of the soft state and into an intermediate/hard state. We exclude this late-time transition from the inference. Its repeated downward offset relative to the soft-state model marks where the fitted thin-disk evolution stops describing the data. The physical interpretation of this transition is discussed in Section~\ref{sec:disk_state}.

Figure~\ref{fig:E3_uvx} shows the posterior-predictive fit to the Epoch~3 \textit{Swift}/XRT light curve (left) and the contemporaneous \textit{HST} UV/optical SED (right) within the direct all-epoch inference. The UV observations used to construct the SED occur within a narrow time window centered at $t\simeq 17.4\pm5$~d after the first X-ray detection included in the fit (vertical shaded band in the left panel).

The UV/optical SED (right panel of Figure~\ref{fig:E3_uvx}) is reproduced by the outer-disk emission of a multi-temperature disk whose spectral peak falls in the soft X-ray band, as constrained by the joint fit. The joint fit must reproduce both the inner-disk luminosity at the UV epoch and the outer-disk spectral normalization and slope. The narrow posterior-predictive band in the UV shows that these data constrain the overall temperature scale of the disk, and therefore the combinations of $(\log_{10} M_{\bullet},\, a_\bullet,\, \cos\iota,\, \log_{10} r_0)$ that determine the relativistic spectrum. The fit reproduces the UV/optical SED using a color-corrected disk spectrum \citep{Done2012MNRAS} without an irradiation component.

\subsection{Constraints on shared and epoch-dependent parameters}
\label{sec:posterior_constraints}

\begin{figure*}[htbp]
\centering
\includegraphics[width=0.8\textwidth]{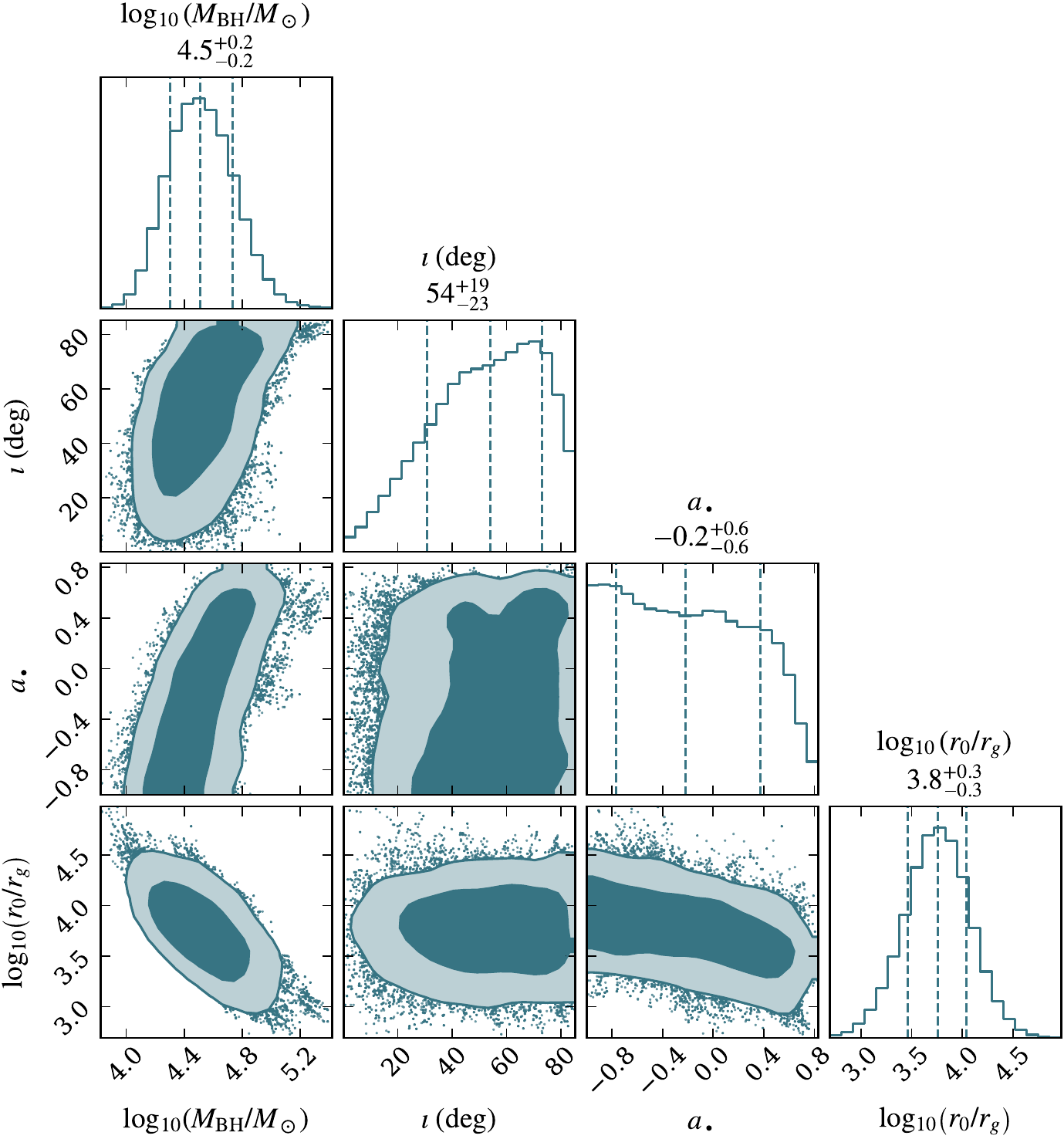}
\caption{
Corner plot of the shared-parameter posterior from the direct all-epoch fit. Shown are the marginalized 1D and 2D projections for $\log_{10}(M_{\bullet}/M_\odot)$, $\iota$, $a_\bullet$, and $\log_{10}(r_0/r_g)$, where $\iota$ is given in degrees.
The contours enclose the 68\% and 95\% highest-posterior-density credible regions, with individual posterior samples shown as points. In each diagonal panel, the dashed vertical lines mark the 16th, 50th, and 84th percentiles, and the title reports the median value and its central 68\% credible interval.
}
\label{fig:shared_corner}
\end{figure*}

\begin{figure*}[htbp]
\centering
\includegraphics[width=\textwidth]{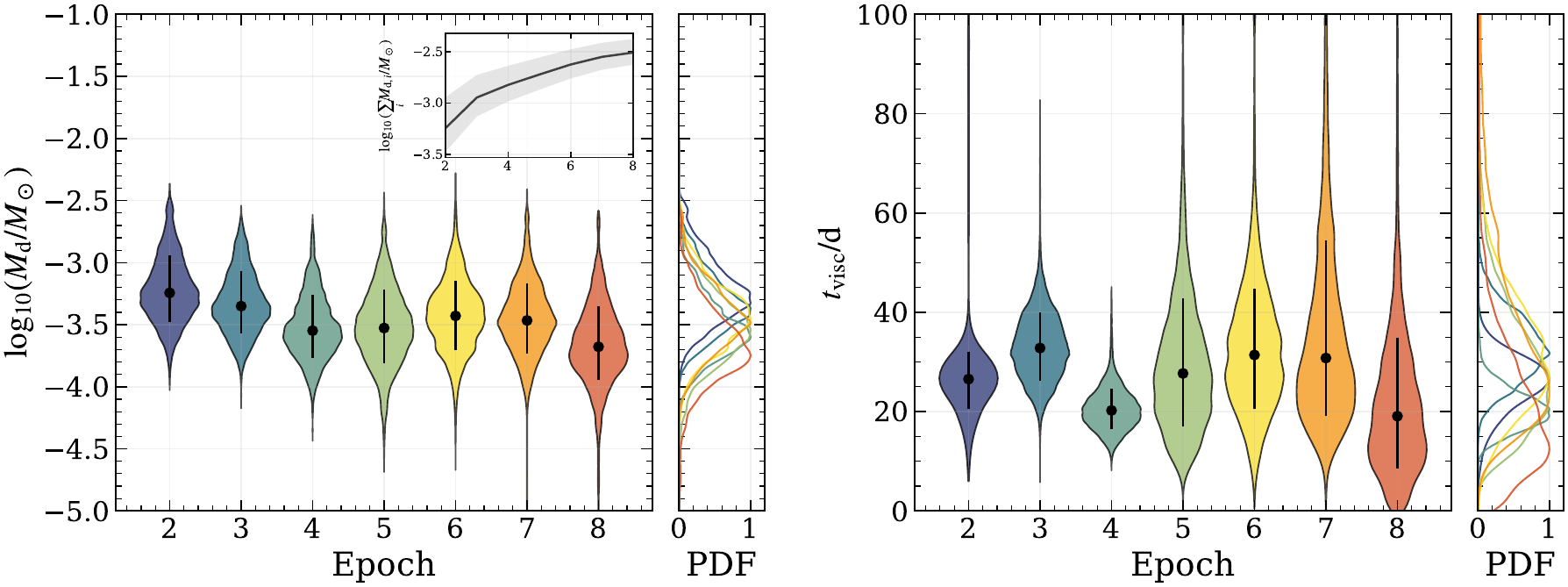}
\caption{
Posterior distributions of the epoch-dependent parameters from the direct all-epoch fit. The left panel shows $\log_{10}(M_{{\rm d},i}/M_\odot)$ and the right panel shows $t_{{\rm visc},i}$ for Epochs~2--8. For each epoch, the colored violin represents the full posterior density, while the black point marks the median and the vertical bar indicates the central 68\% credible interval (16th--84th percentiles). Numerical summaries of all epoch-dependent parameters, including $t_{0,i}$, are given in Table~\ref{tab:epoch_params}.
}
\label{fig:epoch_violin}
\end{figure*}

Figure~\ref{fig:shared_corner} summarizes the marginalized posterior constraints on the shared parameters from the direct all-epoch fit. We infer $\log_{10}(M_{\bullet}/M_\odot)=4.5^{+0.2}_{-0.2}$, $\iota = 54^{+19}_{-23}\deg$, $a_\bullet=-0.2^{+0.6}_{-0.6}$, and $\log_{10}(r_0/r_g)=3.8^{+0.3}_{-0.3}$ (all central 68\% credible intervals). The spin posterior is broad and largely reflects the prior, with only a mild preference against high prograde values. The black hole mass therefore lies in the intermediate-mass regime, as expected \citep[e.g.,][]{Farrell2009Nature,Davis2011ApJ}, while the inferred circularization scale is large in gravitational units, $r_0\sim 10^{3.8}r_g$, where $r_g \equiv GM_{\bullet}/c^2$. For the posterior median mass, this corresponds to $r_g\sim 4.8\times10^{9}$~cm and hence $r_0\sim 2.8\times10^{13}$~cm. For a solar-type donor, the corresponding tidal radius at the posterior median black-hole mass is $R_t \simeq 2.2\times10^{12}$~cm; propagating the full posterior gives $r_0/R_t = 12^{+8}_{-5}$. In other words, $r_0$ is consistent with a partial-TDE circularization scale.

The 2D posterior projections show several degeneracies. The strongest is the anti-correlation between $M_{\bullet}$ and $r_0/r_g$. The observed SED constrains a temperature scale and therefore an absolute radius scale (i.e., in physical units). Increasing $M_{\bullet}$ changes both the gravitational radius and the characteristic disk temperature for a fixed accretion history, and this can be compensated by shifting the angular-momentum scale encoded by $r_0/r_g$. The UV/optical SED provides complementary leverage on the radial temperature profile and thus tightens the joint posterior. By contrast, the posteriors for $\cos \iota$ and $a_\bullet$ remain broader, as expected because inclination and spin primarily affect the spectral normalization and inner-disk temperature structure, respectively, which pure light-curve fitting cannot fully disentangle. Our mass uncertainty is therefore broadened by marginalization over these parameters.

The shared posterior is therefore compact and effectively single-peaked in these key projections.

The epoch-dependent parameters show the expected variability. The viscous timescale varies from epoch to epoch by a factor of $\sim 2$ (with the shortest $t_{\rm visc}$ corresponding to the most rapidly fading outbursts), while the inferred injected mass varies at the sub-dex level.
The combination of a shared angular-momentum scale ($r_0$) and modest variations in $(M_{\rm d}, t_{\rm visc})$ supports
repeated injections of comparable specific angular momentum but with variable mass supply and/or effective viscous response. The formation-time offset $t_{0,i}$ is not shown in Figure~\ref{fig:epoch_violin}, but its numerical constraints are included in Table~\ref{tab:epoch_params}.

\begin{table*}[htbp]
\centering
\caption{Posterior constraints on the epoch-dependent parameters from the direct all-epoch fit. Quoted intervals are the central 68\% credible intervals. The time-offset parameter $t_{0,i}$ is defined relative to the reference time adopted for each epoch in the light-curve analysis.}
\label{tab:epoch_params}
\begin{tabular}{lccc}
\hline
Epoch & $\log_{10}(M_{{\rm d},i}/M_\odot)$ & $t_{{\rm visc},i}$ (d) & $t_{0,i}$ (d) \\
\hline
2 & $-3.24_{-0.23}^{+0.30}$ & $26.54_{-5.87}^{+5.47}$ & $9.25_{-3.37}^{+2.00}$ \\
3 & $-3.35_{-0.22}^{+0.29}$ & $32.83_{-6.50}^{+7.17}$ & $15.81_{-3.29}^{+1.69}$ \\
4 & $-3.55_{-0.22}^{+0.29}$ & $20.27_{-3.79}^{+4.32}$ & $8.71_{-1.57}^{+1.09}$ \\
5 & $-3.53_{-0.29}^{+0.31}$ & $27.70_{-10.77}^{+15.19}$ & $4.18_{-1.30}^{+1.05}$ \\
6 & $-3.43_{-0.28}^{+0.28}$ & $31.39_{-10.81}^{+13.34}$ & $14.86_{-5.12}^{+3.71}$ \\
7 & $-3.46_{-0.27}^{+0.29}$ & $30.80_{-11.72}^{+23.62}$ & $11.09_{-5.62}^{+3.92}$ \\
8 & $-3.68_{-0.27}^{+0.32}$ & $19.07_{-10.45}^{+15.78}$ & $12.81_{-5.99}^{+3.38}$ \\
\hline
\end{tabular}
\end{table*}

\subsection{Derived energetics, mass budget, and orbital interpretation}
\label{sec:derived_quantities}

We use the fitted parameters to compute three derived quantities. First, we calculate the bolometric luminosity $L_{\rm bol}$. Second, we compare the required injected masses with a stripped-mass budget.
Third, we map the shared angular-momentum scale to orbital elements by identifying $r_0$ with the circularization radius $r_{\rm circ}$ and the observed recurrence interval with the orbital period. The possible contribution of a varying disk-formation delay to the recurrence intervals is discussed separately in Section~\ref{sec:disk_timing}.

\begin{figure*}
    \centering
    \includegraphics[width=\textwidth]{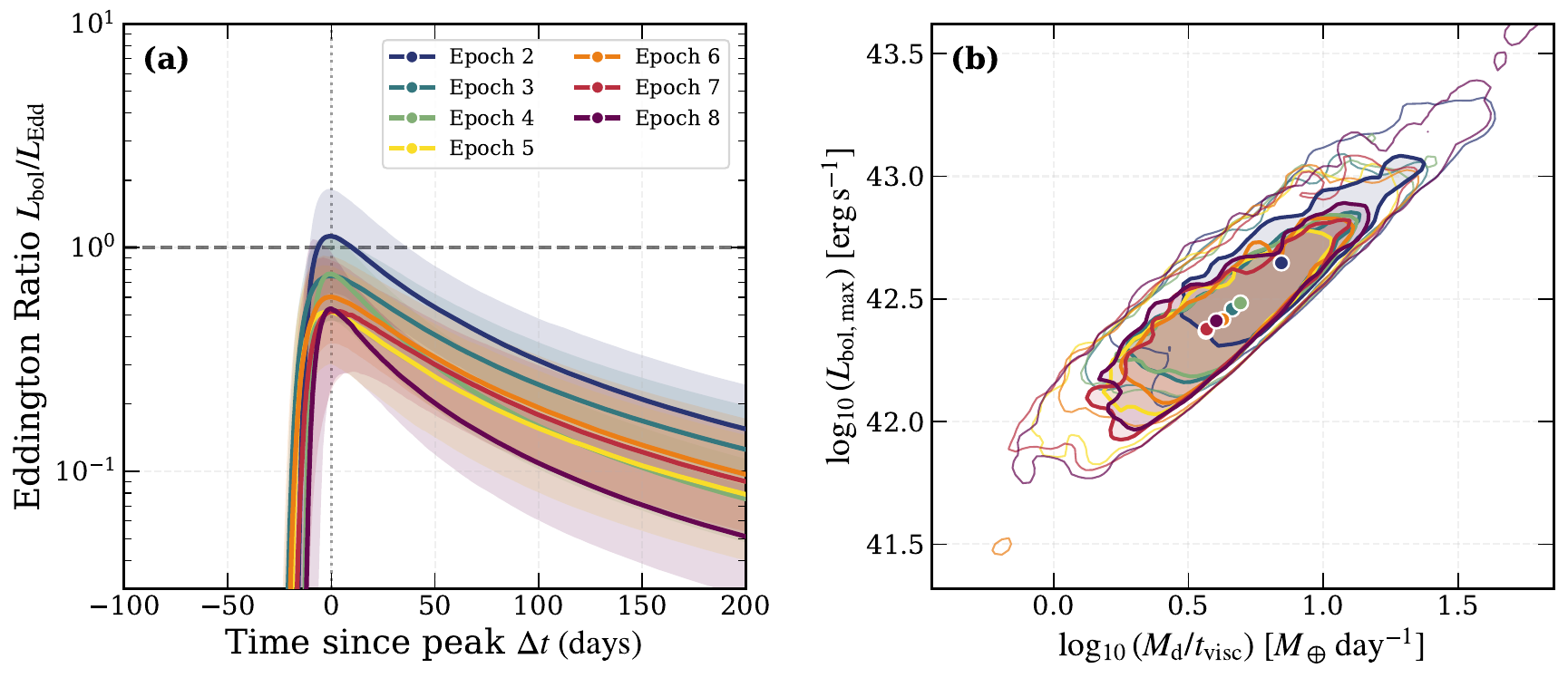}
    \caption{Bolometric luminosity and mass accretion properties across the observed outburst epochs. \textbf{(a)} Evolution of the Eddington ratio ($L_{\rm bol}/L_{\rm Edd}$) relative to the time of peak luminosity ($\Delta t = t - t_{\rm peak} = 0$). Solid curves and shaded regions represent the median model predictions and their 68\% credible intervals, respectively. The horizontal dashed line marks the Eddington limit ($L_{\rm bol} = L_{\rm Edd}$). \textbf{(b)} Joint posterior constraints on the peak bolometric luminosity ($L_{\rm bol,max}$) and the characteristic accretion-rate scale ($M_{\rm d}/t_{\rm visc}$). The accretion rate is expressed in units of Earth masses per day ($M_\oplus\,{\rm day}^{-1}$). Contours denote the 68\% and 95\% credible levels, with colored circles indicating the median values for each epoch. The shared legend indicates the color coding for the successive return epochs.}
    \label{fig:bol_edd}
\end{figure*}

To connect the fitted parameters more directly to the energetics, Figure~\ref{fig:bol_edd} shows the model-inferred bolometric Eddington ratio, $L_{\rm bol}/L_{\rm Edd}$, as a function of time relative to the light-curve peak. All epochs reach their maximum $L_{\rm bol}/L_{\rm Edd}$ close to peak and then decline smoothly over $\sim 10^2$~days. For most of the fitted soft-state interval the posterior medians remain sub-Eddington, but the peak values are generally close to the Eddington limit, and in some cases can reach or slightly exceed it (Epoch~2). The inferred peak luminosities approach the regime where radiation pressure, advection, and outflows can affect the thin-disk description. Excluding the measurements within $\pm10$~d of each peak shifts the shared $M_{\bullet}$ and $r_0$ by $0.22\sigma$ and $0.11\sigma$ respectively, while one viscous timescale shifts more substantially. The Eddington ratio at the end of the fitted soft-state interval, where the source begins its transition to the hard state, spans approximately $0.16$--$0.33$ across epochs, with a typical value of $\sim0.2$. It also mirrors a key observational fact emphasized in long-term \textit{Swift} monitoring: the outbursts evolve in duration and waiting time, while their peak luminosities are comparatively stable along the sequence \citep{Yan2015ApJ}.

The right panel of Figure~\ref{fig:bol_edd} shows how the fitted disk parameters set the inferred bolometric peak luminosity. In viscously spreading ring solutions, the accretion power is controlled by a finite injected mass reservoir $M_{\rm d}$ processed on a viscous time $t_{\rm visc}$. A useful way to express this is
\begin{equation}
\dot M_{\rm in}(t) \;=\; \frac{M_{\rm d}}{t_{\rm visc}}\,
\mathcal{G}\!\left(\frac{t+t_0}{t_{\rm visc}}; a_\bullet, r_0\right),
\end{equation}
where $\mathcal{G}$ is a dimensionless function fixed by the Green's-function evolution and relativistic disk dynamics \citep[e.g.,][]{MummeryBalbus2020}. The characteristic accretion-rate scale is therefore set by $M_{\rm d}/t_{\rm visc}$, with larger values producing larger $L_{\rm bol,max}$. Here $M_{\rm d}$ is the mass initially placed into the viscously evolving disk and $\dot M_{\rm in}(t)$ is the accretion rate its viscous evolution generates; neither is the stellar fallback rate. All of $M_{\rm d}$ is accreted as $t/t_{\rm visc}\rightarrow\infty$, but only a fraction within the fitted interval. For $M_{\bullet}\sim10^{4.5}\,M_{\odot}$ the Eddington accretion rate is small, so $M_{\rm d}\sim10^{-3}\,M_{\odot}$ processed over a few tens of days gives $\dot M_{\rm in}\sim\dot M_{\rm Edd}$ at the inner boundary.

\begin{figure}[htbp]
\centering
\includegraphics[width=\columnwidth]{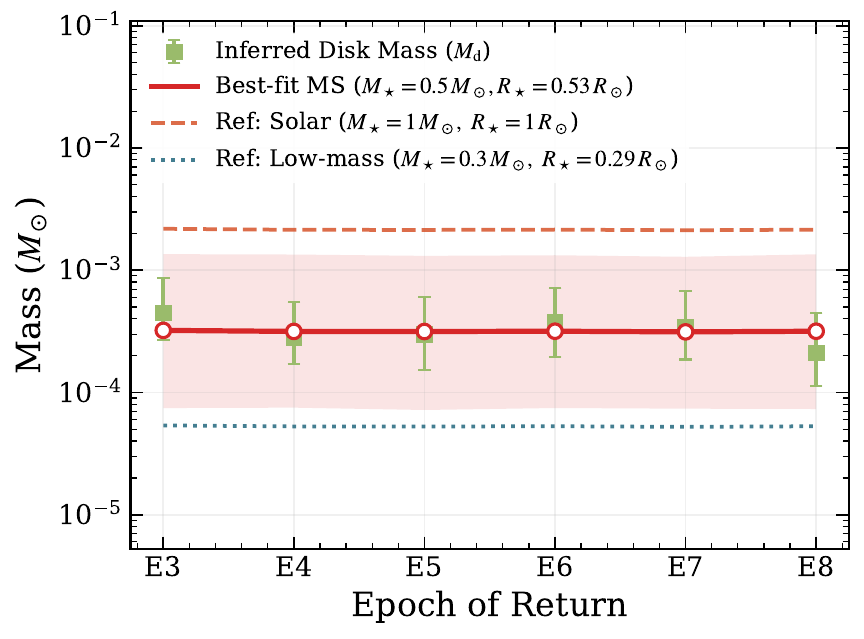}
\caption{Comparison of inferred injected mass per outburst ($M_{\rm d}$, squares) and theoretical stripped masses ($M_{\rm strip}$, lines) for each epoch. Stripped masses are estimated via $M_{\rm strip} \approx \frac{1}{2}\beta^3 M_\star$ \citep{Ryu2020_pTDE}. Reference models for a solar-type (dashed) and a low-mass dwarf (dotted) are shown for context. The solid red line (with 68\% CI shaded) represents the best-fitting main-sequence donor derived by matching $M_{\rm strip}$ to $M_{\rm d}$ assuming $R_\star \propto M_\star^{0.9}$.}
\label{fig:mass_budget}
\end{figure}

We next test whether the inferred injected mass per outburst can be supplied by repeated partial stripping.
Figure~\ref{fig:mass_budget} shows that the inferred injected masses cluster at $M_{\rm d}\sim 10^{-3}\,M_\odot$ across Epochs~3--8, with only modest scatter. In a repeating partial-TDE picture, this injected mass is naturally identified with the bound debris mass circularized per pericenter passage \citep[recalling that in tidal encounters roughly half of the removed mass is typically bound, e.g.,][]{Ryu2020_pTDE}. The figure compares these inferred $M_{\rm d}$ values to stripped-mass expectations parameterized in terms of the penetration factor $\beta\equiv R_t/r_p$, where $R_t\simeq R_\star (M_{\bullet}/M_\star)^{1/3}$ is the tidal radius. For each posterior sample, we identify $r_0$ with a circularization radius $r_{\rm circ}$ and adopt $r_p\simeq r_{\rm circ}/2\simeq r_0/2$, the high-eccentricity limit of $r_{\rm circ}=(1+e)r_p$, and then compute $\beta=R_t/r_p$ for each assumed donor mass and radius. Hydrodynamical simulations show that the mass lost in partial disruptions depends steeply on encounter strength and, importantly, also on stellar structure \citep[often expressed through polytropic index or realistic stellar models, e.g.,][]{GuillochonRamirezRuiz2013,Ryu2020_pTDE}.
We adopt the approximate scaling of \citet{Ryu2020_pTDE},
\begin{equation}
M_{\rm strip} \approx \frac{1}{2}\,\beta^3\,M_\star,
\end{equation}
where $M_{\rm strip}$ denotes the bound stripped mass available to form the disk, which follows from the derived $\beta$ and is compared directly with $M_{\rm d}$. This relation fixes the structure-dependent normalization, often parameterized by the critical penetration factor $\beta_c$ (here $\beta_c=1$). With this parametrization, a single low-mass main-sequence donor can account for the disk-mass scale inferred across the modeled epochs: the ``best-fitting'' donor indicated in Figure~\ref{fig:mass_budget} has $M_\star\simeq 0.5\,M_\odot$ and $R_\star\simeq 0.53\,R_\odot$, and yields stripped masses of order $10^{-3}\,M_\odot$ per passage, consistent with the injected masses required by the light-curve fits. This comparison links the inferred outburst energetics to a stellar fuel reservoir without invoking fine-tuned radiative efficiencies or an additional long-lived mass store.

\begin{figure*}[htbp]
\centering
\includegraphics[width=0.8\textwidth]{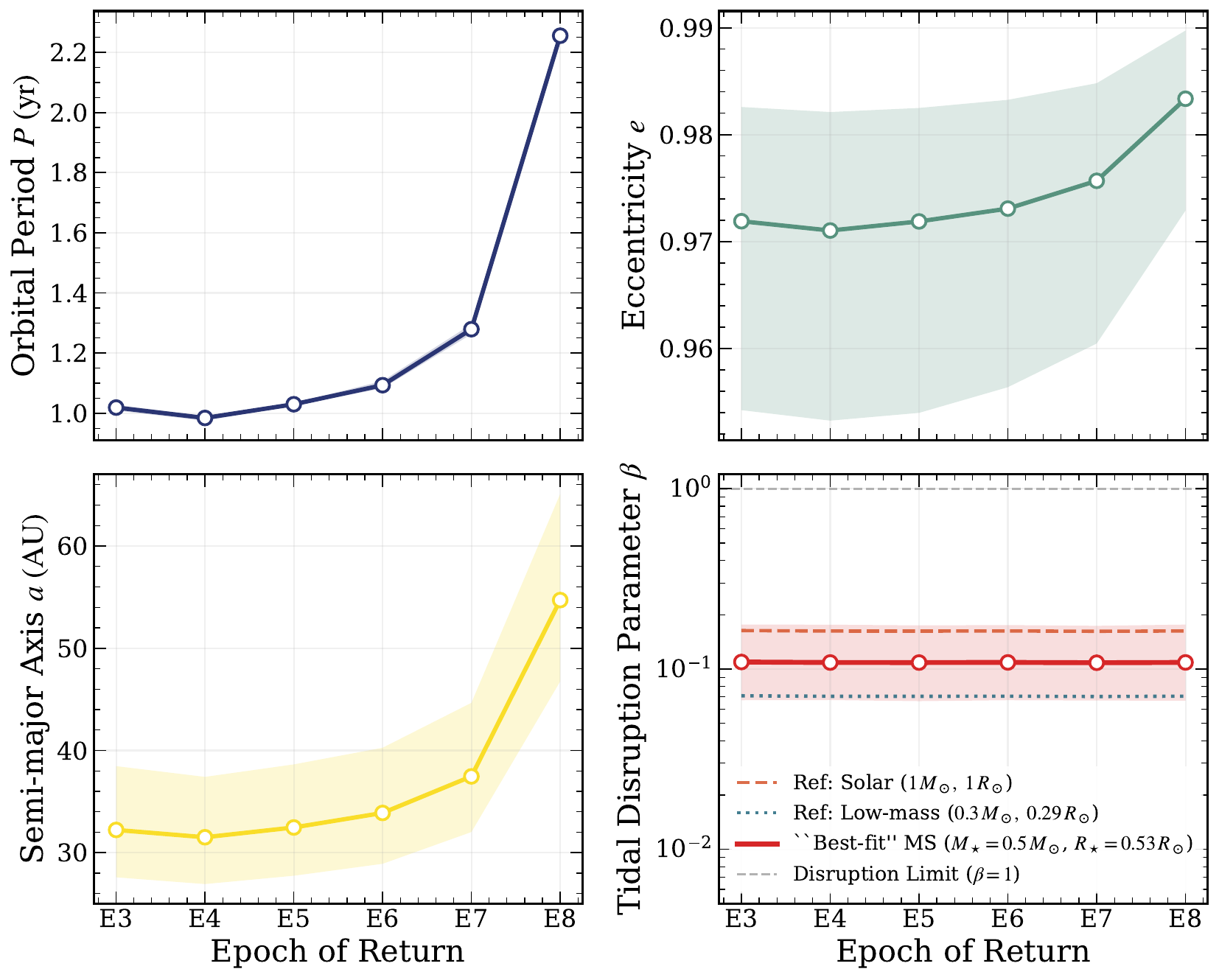}
\caption{Evolution of the derived orbital parameters across successive outburst epochs (E3--E8). The panels display the orbital period $P$ (top left), eccentricity $e$ (top right), semi-major axis $a$ (bottom left), and the tidal disruption parameter $\beta \equiv R_t/r_p$ (bottom right). Note that $P$ is the observed recurrence interval interpreted as an orbital period, without correcting for the disk formation offset (i.e., the delay between pericenter passage and luminous disk formation; see Section~\ref{sec:disk_timing}). Open circles and shaded regions denote the median values and 68\% credible intervals derived from the MCMC posterior distributions. The pericenter distance $r_p$ is calculated by identifying the fitted characteristic radius with a circularization radius, $r_0\simeq r_{\rm circ}$, and using the Keplerian relation $r_{\rm circ}=(1+e)r_p$ (which yields $r_p\approx0.5r_0$ for highly eccentric orbits). In the bottom-right panel, the solid line and shaded band represent $\beta$ evaluated for the best-fitting main-sequence donor star derived from our mass budget analysis (see Figure~\ref{fig:mass_budget}). For context, expected $\beta$ values for a solar-type star ($1\,M_\odot, 1\,R_\odot$; dashed line) and a low-mass dwarf ($0.3\,M_\odot, 0.29\,R_\odot$; dotted line) are also shown as references. The gray dashed horizontal line indicates the full disruption limit ($\beta = 1$).}
\label{fig:orbital_params}
\end{figure*}

We translate the inferred angular-momentum scale into orbital elements using the mapping adopted above. For a bound Keplerian orbit,
\begin{equation}
r_{\rm circ} = (1+e)\,r_p.
\end{equation}
The returning debris is expected to span a distribution of specific angular momenta,
while the fitted narrow-ring radius represents a characteristic circularization scale. We therefore identify $r_0$ with $r_{\rm circ}$ when deriving $r_p$, $e$, and $\beta$. For highly eccentric orbits ($e\rightarrow 1$), this gives $r_p\simeq r_{\rm circ}/2\simeq r_0/2$ and a semi-major axis $a=r_p/(1-e)\simeq r_0/(1-e^2)$, so even small changes in $e$ can produce large changes in $a$ and hence in the orbital period $P\propto a^{3/2}$. Under this mapping, the inferred orbital parameters are consistent with a repeating, grazing tidal-encounter picture: the eccentricity remains high, the semi-major axis increases from tens of AU to somewhat larger values, and the period inferred from the earlier returns remains of order a year. The penetration factors inferred for the best-fitting donor remain safely below the full-disruption limit, $\beta\simeq 0.08$--$0.11$ across E3--E8, consistent with repeated partial stripping.

Figures~\ref{fig:bol_edd}--\ref{fig:orbital_params}
connect the fitted disk solution to three physical properties of the system. First, the fitted parameters give near-Eddington bolometric peaks during the soft-state phase and show the expected amplitude--timescale covariance of a viscously draining disk (Figure~\ref{fig:bol_edd}). Second, the required injected masses per outburst, $M_{\rm d}\sim10^{-3}\,M_\odot$, can be supplied by partial stripping of a single low-mass main-sequence donor, so the inferred fuel requirement can be accounted for without invoking an additional long-lived reservoir (Figure~\ref{fig:mass_budget}).
Third, the shared angular-momentum scale $r_0$ maps to a highly eccentric, grazing orbit with $\beta<1$ and year-scale recurrence (Figure~\ref{fig:orbital_params}).

We next discuss the repeating partial-disruption interpretation.

\section{Discussion}
\label{sec:discussion}

The multi-epoch fit identifies a repeatable injected-mass scale and outburst-to-outburst diversity that is captured mainly by changes in the injected mass and viscous timescale. We discuss these results in the context of repeating partial tidal disruption, then address the soft-to-hard state transition, the recurrence-time behavior, and alternative interpretations.

\subsection{Support for a repeating partial disruption interpretation}
\label{sec:disk_ptde}

These results support repeating partial tidal disruption (rpTDE), which has recently gained both observational and theoretical support as a channel for repeating accretion transients \citep[e.g.,][]{Rossi2021SSRv,Gezari2021ARAA, Lin2024AT2022dbl,Bandopadhyay2024ApJ,Somalwar2025AT2020vdq}. The joint fit yields a characteristic disk-formation radius shared across the modeled sequence, and the free-$r_0$ analysis below tests this shared-scale description directly. Identifying $r_0$ with the circularization scale makes it an angular-momentum diagnostic: for an orbit with specific angular momentum $j$ around the black hole, the circular orbit corresponding to that angular momentum is at a radius
\begin{equation}
r_{\rm circ} \equiv \frac{j^2}{G M_{\bullet}}.
\end{equation}
For bound orbits
\begin{equation}
r_{\rm circ} = (1+e)\,r_p,
\label{eq:rcirc_e_rp}
\end{equation}
which is the mapping we adopt when translating $r_0$ to orbital elements. Within this mapping, a nearly constant $r_0$ implies a nearly constant characteristic $j$, and hence a repeatable pericenter geometry, even if the orbital energy and return time evolve.

\begin{figure}[htbp]
\centering
\includegraphics[width=\columnwidth]{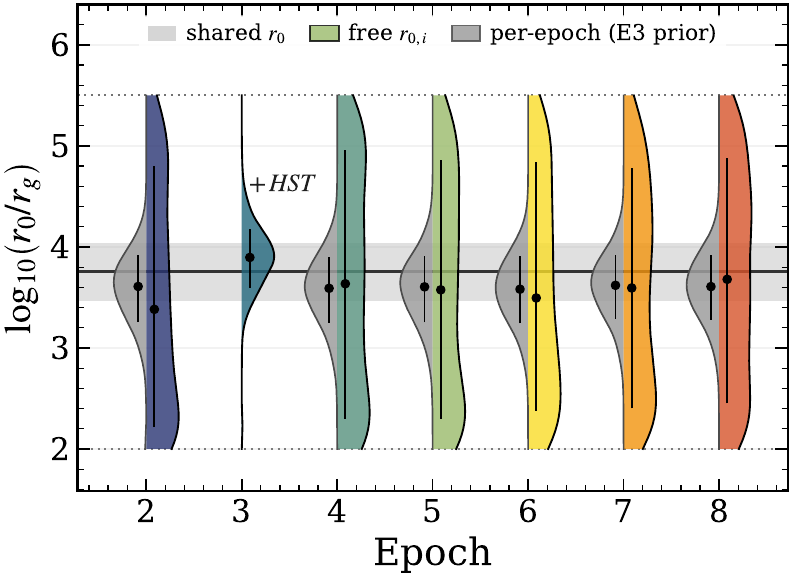}
\caption{Three treatments of the disk-formation radius $r_0$, compared epoch by epoch. The grey band and solid line show the 68\% credible interval and median of the shared $r_0$ from the fiducial all-epoch fit. For each epoch the left (grey) half-violin is the posterior of an independent per-epoch fit that adopts one-dimensional kernel density estimates of the Epoch~3 marginals as priors on the global parameters; Epoch~3 has no grey half-violin because it defines those priors. The right half-violin, colored by epoch, is the posterior of the joint all-epoch fit in which $r_0$ is instead free for every outburst, with a broad top-hat prior $\log_{10}(r_0/r_g)\in[2.0,5.5]$; the dotted lines mark that prior range, and the violins terminate on it. Black points and bars give the median and 16th--84th percentiles. Epoch~3 is the only epoch with a contemporaneous \textit{HST} SED, as marked.}
\label{fig:r0_violin}
\end{figure}

We repeated the all-epoch fit with $r_0$ free for every outburst under a broad top-hat prior. Only Epoch~3, which includes the contemporaneous \textit{HST} SED, yields an informative individual constraint; the X-ray-only epochs remain largely prior dominated (Figure~\ref{fig:r0_violin}). The free-$r_0$ posteriors provide no evidence for a systematic trend across the sequence, and the remaining shared parameters are unchanged.

A shared $r_0$ is expected if the same donor repeatedly grazes the black hole on a bound orbit. Hydrodynamic simulations of partial disruptions show that asymmetric mass loss can alter the remnant orbit while leaving the angular momentum less strongly affected than the energy, with the details depending on encounter depth and stellar structure \citep[e.g.,][]{Bonnerot2016MNRAS,Ryu2020_pTDE,Chen2024ApJ}. Thus, the remnant orbit can evolve while its angular momentum, and hence $r_0$ stays nearly fixed.

The inferred injected masses are also compatible with the mass loss expected from repeated partial stripping. The fitted values are of order $10^{-3}\,M_\odot$ per outburst: much smaller than a full disruption, but large enough to require a stellar fuel reservoir. This is the expected regime for repeated partial stripping of a surviving donor. The modest epoch-to-epoch scatter in $M_{\rm d}$ can be attributed to variations in encounter depth, circularization efficiency, or the evolving stellar structure. Multi-encounter simulations show that repeated tidal heating and mass loss can progressively modify the donor and its response to later passages \citep{Sharma2024MNRAS,Liu2025ApJ}.

These results support an rpTDE interpretation of the fitted soft-state evolution: the same physical picture accommodates the shared angular-momentum scale, the repeated fuel requirement, and the viscous light-curve diversity. The recurrence-time history is discussed further in Section~\ref{sec:disk_timing}.

\subsection{State transitions and the terminal rapid decline}
\label{sec:disk_state}

The fitted soft-state evolution is often followed by a much faster decline, accompanied by spectral hardening in the HR curve. The fitted interval therefore corresponds to the soft-state portion of each outburst, while the excluded late-time measurements trace the subsequent transition toward an intermediate/hard state.

As the disk drains, the source may eventually leave the radiatively efficient thin-disk regime that describes the fitted light-curve body. In that case, the emergent spectrum is expected to harden, while the observed 0.3--10 keV light curve need not track the bolometric output in a simple one-to-one way. This matters for \textit{Swift}/XRT, whose effective area is strongly energy dependent \citep{Burrows2005SSRv}, and for which our conversion from count rate to flux depends on the source spectrum and its evolution \citep{Evans2010AA}. As a result, if the source spectrum hardens substantially during the late-time transition, the observed 0.3--10 keV count rate and any band-limited luminosity inferred under a fixed or time-averaged spectral assumption can decline more abruptly than the underlying accretion power.

The terminal drop is not evidence for a sudden disappearance of the accretion flow. Rather, it marks a rapid soft-state exit, and the physical interpretation is that the source is transitioning into an intermediate/hard state in which the thin-disk model used for the fitting is no longer adequate. Allowing $r_0$ to vary between epochs does not alter this picture: refitting with the spectrally hardened measurements included and $r_0$ free per epoch shortens the inferred viscous timescales but degrades the description of the soft state, and does not reproduce the sharp late-time decline.

The late-time drop should also be separated from the recurrence mechanism itself. We associate the recurrence with episodic delivery and viscous processing of new material, while the terminal decline marks the end of the radiatively efficient soft-state phase of a given outburst. This behavior is analogous to the soft-to-hard state transitions observed in Galactic X-ray binaries, where the transition occurs as the accretion rate drops below a critical Eddington ratio of order a few per cent \citep[e.g.,][]{Fender2004MNRAS,Remillard2006ARAA,Done2007AARv}. In HLX--1 the transition appears to occur at a somewhat higher ratio, very roughly $L_{\rm bol}/L_{\rm Edd}\sim 0.2$ (Figure~\ref{fig:bol_edd}), perhaps a factor few above the canonical XRB value. A similarly soft-to-hard transition luminosity (of order a few percent of Eddington) has been found in the broader TDE population \citep{GoodwinMummery2026arXiv}, and so HLX--1 is somewhat of an outlier within the observed accretion disk population.

In the canonical picture this disk transition implies that a geometrically thick, radiatively inefficient flow \citep{Narayan&Yi1995ApJ,YuanNarayan2014} has formed within the disk's inner regions accompanied, in some systems, by the launch of a compact jet \citep{Fender2004MNRAS}. The jet activity already documented in HLX--1 \citep{Webb2012Science} is consistent with this picture. A full end-to-end description of HLX--1 beyond the soft state would require coupling the fitted thin-disk evolution to a model for the subsequent state transition and for the spectral response of the source as it hardens.

\subsection{Orbital timing and the recurrence-time puzzle}
\label{sec:disk_timing}

The recurrence history of HLX--1 contains information beyond the shared angular-momentum scale alone. The timing phenomenology contains two distinct puzzles. The first is the modest but systematic increase in the interval between the earlier outbursts, which may resemble a positive secular drift in the recurrence interval. The second is the much more dramatic change between E7 and E8, where the interval increases to nearly twice the earlier value, after which no further major outburst is seen.

The observed interval between successive luminous flares is not necessarily the orbital period. If the flare associated with the $n$th return occurs at
\begin{equation}
t_{{\rm flare},n} = t_{{\rm peri},n} + \tau_{{\rm form},n},
\end{equation}
where $\tau_{{\rm form},n}$ denotes the delay between pericenter passage and the formation of the luminous disk, then the observed interval is
\begin{equation}
\Delta t_{{\rm flare},n}=t_{{\rm flare},n+1}-t_{{\rm flare},n}=P_{{\rm orb},n}+\left(\tau_{{\rm form},n+1}-\tau_{{\rm form},n}\right).
\end{equation}
The orbital mapping in Section~\ref{sec:derived_quantities} assumes that the disk-formation/circularization delay does not vary strongly across epochs. The disruption time is not directly observed, so any delay between pericenter passage and luminous disk formation, whether from weak relativistic precession, slow stream self-intersection, or inefficient circularization, is absorbed into $\tau_{\rm form}$. If $\tau_{\rm form}$ varies between returns, changes in the observed recurrence interval need not directly trace changes in the orbital period.

For the earlier outbursts, this ambiguity matters. The gradual increase in interval through the first several returns is modest enough that it could reflect small changes in disk-formation or circularization time rather than a large secular drift in the orbit itself. This avoids requiring a significant positive $\dot P$ at every passage; in a repeating stripping scenario, the time required for debris to self-intersect, dissipate orbital energy, and form the luminous disk need not be strictly constant from one encounter to the next, especially if the stellar structure, debris geometry, or stream self-intersection conditions evolve slowly.

If, however, one interprets the early interval increase as a pure orbital effect, then a positive interval derivative corresponds to a lengthening orbital period and therefore to a net gain in orbital energy by the surviving star. The rpTDE literature does contain mechanisms that can alter the period, but the sign and magnitude depend strongly on the encounter regime. For example, \citet{Bandopadhyay2024ApJ} showed that tides can in some cases produce appreciable period evolution in repeating partial-disruption systems, while \citet{Linial2024MNRAS} emphasized that repeated star--disk interaction and cumulative tidal effects can shape the long-term timing behavior.

At the same time, not every route to period evolution is compatible with the shallow-grazing regime inferred under this mapping. In particular, \citet{Cufari2023MNRAS} showed that positive orbital-energy kicks occur most readily in much deeper partial disruptions, where asymmetric mass loss can eventually eject the surviving core rather than leave it on a long-lived bound orbit. This is qualitatively different from the regime suggested by our inferred $\beta$, which is much smaller and instead places HLX--1 in a much shallower encounter regime. More recently, \citet{Bandopadhyay2025ApJ} argued that tidal heating itself is not expected to provide an arbitrarily large energy reservoir for the surviving star. Attributing the early interval increase to a large positive orbital-energy change would therefore be in tension with the shallow-encounter regime inferred above.

\begin{figure}[htbp]
\centering
\includegraphics[width=\columnwidth]{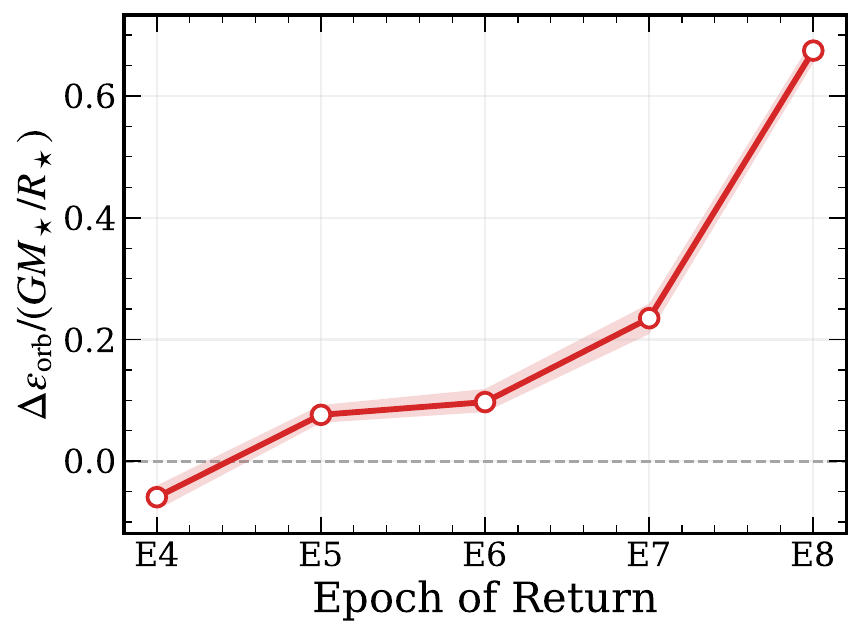}
\caption{Inferred orbit-to-orbit change in specific orbital energy, normalized by $GM_\star/R_\star$ for the best-fitting donor from Figure~\ref{fig:mass_budget}, under the assumption that the observed interval changes are produced entirely by orbital-period evolution. Open circles show the median values and the shaded band the 68\% credible interval.}
\label{fig:orbital_energy_change}
\end{figure}

This point can be quantified more directly. If the observed interval changes are interpreted entirely as orbital-period evolution, then each step in the recurrence sequence implies a corresponding change in the donor's specific orbital energy,
\begin{equation}
\Delta \epsilon_{{\rm orb},n}
\equiv
\epsilon_{{\rm orb},n+1}-\epsilon_{{\rm orb},n},
\end{equation}
where for a Keplerian orbit $\epsilon_{\rm orb}=-GM_{\bullet}/(2a)$ and $P_{\rm orb}\propto a^{3/2}$. Figure~\ref{fig:orbital_energy_change} shows the resulting $\Delta \epsilon_{\rm orb}$ values normalized by the stellar binding-energy scale $GM_\star/R_\star$, evaluated using the ``best-fitting'' donor from the mass-budget analysis of Figure~\ref{fig:mass_budget}.

The required energy change grows gradually from E4 to E7, but the final step to E8 is dramatically larger, approaching a substantial fraction of the stellar binding energy. The last jump is therefore already uncomfortably large for any mechanism based on gentle secular evolution. This is difficult to reconcile with either shallow partial stripping or tidal heating: if the orbit were to gain an energy comparable to the stellar binding energy on a single return, one would expect the encounter itself to be highly destructive. The late-time timing behavior appears too extreme to be explained by a simple smooth positive $\dot P$ picture.

The E7--E8 interval therefore need not correspond to a single orbital period. Either the orbit genuinely expanded, or an intervening return stripped too little mass or produced debris that failed to circularize into a luminous disk, yielding an apparent doubling of the flare interval.
The system may then have undergone a pericenter passage without generating an outburst of the same phenomenological class as the earlier events. There is no evidence that a full major flare was simply missed, but the data near the nominal expected return are not of sufficient quality to exclude all weaker or shorter-lived activity \citep[e.g.,][]{Yan2015ApJ}. Changes in the stripped mass as the surviving donor evolves, in the debris geometry, or in the circularization efficiency can each suppress a luminous disk-forming episode.

An isolated failure of disk formation would nevertheless represent a departure from the otherwise stable outburst phenomenology. \textit{Swift}/XRT follow-up through 2026 shows no subsequent major X-ray outburst after E8. Thus, the E7--E8 jump does not appear to be followed by a return to the earlier recurrent behavior. The large E7--E8 interval and the absence of later major flares instead suggest a qualitative change in the system.
Either cumulative positive energy changes left the surviving donor unbound, ending further returns, or the donor survived on a wider orbit where stripping or luminous disk formation became too inefficient to power a comparable outburst. The present data do not distinguish between these outcomes.

HLX--1 is not alone in presenting unexplained recurrence-time evolution. ASASSN-14ko, the best-established repeating nuclear transient, shows a smoothly decreasing period with $\dot P = -0.0026\pm0.0006$ \citep{Payne2023ApJ}. \citet{Linial2024MNRAS} argued that the most plausible mechanism is hydrodynamical drag through an accretion disk, but this requires a disk more massive than the star. The repeating partial-TDE candidate eRASSt~J045650.3$-$203750 exhibits a rapid decrease in recurrence time from $\sim 300$ to $\sim 190$~d, which hydrodynamic simulations can reproduce only if the donor loses $\sim 80$--$90\%$ of its mass in the initial encounter \citep{Liu2024AA}. Most recently, the quasi-periodic eruption source ZTF19acnskyy (``Ansky'') was found to have a positive period derivative, $\dot P \approx 1.7\times10^{-2}$ \citep{Chakraborty2026ApJL}, opposite in sign to predictions from inspiral or gas-drag models and unexplained by any mechanism the authors tested. Whether the recurrence time increases (HLX--1, Ansky) or decreases (ASASSN-14ko, J0456$-$20), reproducing the secular timing behavior requires fine-tuned or physically extreme conditions, so no single mechanism currently provides a general explanation for the recurrence-time evolution in these systems.

Modest variations in the disk-formation delay may account for the early interval drift, whereas the E7--E8 transition requires a qualitative change in the orbital or disk-formation behavior.
\subsection{Relation to existing interpretations of the HLX--1 outburst cycle}
\label{sec:disk_lit}

HLX--1 has been discussed within several broad classes of models, which differ mainly in what sets the recurrence and in whether the accretor must be an intermediate-mass black hole (IMBH). The IMBH interpretation is motivated by the extreme luminosity and disk-dominated X-ray states \citep[e.g.,][]{Farrell2009Nature,Davis2011ApJ}. Long-term \textit{Swift} monitoring established a quasi-recurrent sequence with an increasing waiting time and evolving outburst duration \citep{Yan2015ApJ}. The present fit adds cross-epoch constraints on a repeated injected-mass scale, a characteristic angular-momentum scale, and epoch-to-epoch changes in the mass and viscous response.

Disk-instability models attribute the outbursts to accretion-disk instabilities analogous to those seen in Galactic X-ray transients, most commonly the thermal--viscous disk instability model (DIM) \citep[e.g.,][]{Lasota2001}. For HLX--1, however, the combination of long recurrence times and comparatively short outburst durations has long been difficult to reconcile with a straightforward DIM trigger at the established distance \citep{Lasota2011ApJ}. In practice, instability-based scenarios tend to require additional ingredients, such as irradiation, truncation, or externally modulated supply, to reproduce the observed timescales. Our results sharpen that tension: a successful model must account not only for the recurrence and duration, but also for the large disk scale constrained by the Epoch~3 broadband data and the repeated requirement of a $\sim$ constant injected mass reservoir. These features arise naturally in an episodically fueled disk, but not generically in a global instability cycle.

Feedback-regulated models instead invoke inner-flow limit cycles, such as wind- or irradiation-driven oscillations \citep[e.g.,][]{Soria2017MNRAS} or radiation-pressure-driven instabilities \citep[e.g.,][]{Wu2016ApJ}. Such mechanisms can reproduce sub-Eddington peaks and state changes, and in principle can be tuned to yield year-like timescales. Their main difficulty is different: both the recurrence and the amplitude depend on uncertain feedback efficiencies and on how wind launching, irradiation geometry, and accretion rate couple to one another. More importantly, these models do not by themselves explain the repeated requirement of a modest but well-defined mass reservoir of order $10^{-3}\,M_\odot$ per event. In a repeating partial-disruption picture, that fuel scale and the stream angular-momentum scale arise from the same orbital encounter. This point is reinforced by the UV/optical spectrum: our color-corrected disk model reproduces the Epoch~3 broadband SED without an irradiation component (Section~\ref{sec:ppd_fits}), reducing the need for irradiation in this particular SED fit. The recurrence in the repeating partial-disruption picture is instead set by the orbital return time, so neither irradiation feedback nor a global disk instability is required to regulate the outburst cycle.

Geometric models interpret the $\sim$year timescale as a superorbital modulation produced by geometric effects such as precession and beaming in a super-Eddington stellar-mass system, by analogy with SS433-like sources \citep{King2014MNRAS}. Purely geometric modulation can in principle generate recurring bright phases without requiring large changes in the mass supply. The challenge is that HLX--1 is constrained by more than recurrence alone: the source also shows a thermal spectral component, a multiwavelength spectrum requiring a large disk scale, and an outburst sequence whose peak energetics are organized as expected for viscous draining of a finite reservoir. A purely orientation-driven scenario does not predict those fitted features without additional assumptions.

More closely related are models with episodic mass supply from a donor on an eccentric orbit, with transfer enhanced at periastron through tidal interaction or mass transfer \citep[e.g.,][]{Lasota2011ApJ,vanderHelm2016MNRAS}. These already contain two ingredients that are also central to our interpretation: an external fueling clock and a disk response time.
The repeating partial-disruption picture is a physically specific realization of this broader family. It links the shared $r_0$ to the stream angular-momentum scale, the required fuel per outburst to the stripped-mass budget, and the recurrence sequence to the same orbital and disk-formation framework.

\subsection{Caveats}
\label{sec:disk_caveats}

The analysis has several limitations. First, the model assumes a geometrically thin, radiatively efficient disk and is applied only to the soft-state portion of each outburst. The fitted peaks approach the Eddington luminosity, where radiation pressure, advection, and outflows may become important \citep[e.g.,][]{Huang2023ApJ,Huang2024ApJ,Price2024,Sharma2026}; the peak-exclusion test shows that the shared parameters are robust, although individual viscous timescales are more sensitive. If an outflow removes stripped debris before it joins the disk, the fitted $M_{\rm d}$ is a lower bound on $M_{\rm strip}$.

Second, the likelihood treats each epoch's X-ray emission as arising from a single newly injected ring. Residual emission from earlier disks contributes $\sim 15$--$20\%$ of each epoch's peak X-ray luminosity; because the current outburst dominates, the shared parameters are minimally affected, but the approximation is less secure for the UV/optical emission, where earlier disks cool more slowly. Paper~II adopts stacking in the likelihood itself.

Third, the mapping from $r_0$ to $(P,e,r_p)$ is model-dependent. We identify $r_0$ with the characteristic circularization scale, but the debris may span a distribution of specific angular momenta and undergo angular-momentum redistribution during disk formation. The recurrence interval may also differ from the orbital period if the disk-formation delay varies between epochs (Section~\ref{sec:disk_timing}). Alternative circularization and stellar-structure-dependent stripping prescriptions would quantitatively shift the inferred orbital elements, donor properties, and encounter depths.

Fourth, the present analysis uses only the \textit{HST} epoch contemporaneous with the soft X-ray state as a broadband constraint on the disk model. Only Epoch~3 provides such contemporaneous coverage, and the disk-only model self-consistently accounts for its emission from the UV/optical/IR bands to X-rays. At other \textit{HST} epochs without contemporaneous soft X-ray coverage, the outer disk can remain luminous after the hotter inner disk has faded because it cools more slowly \citep[e.g.,][]{vanVelzen19,MummeryBalbus2020}. The broader multi-epoch UV/optical/IR data set may therefore contain contributions from both the evolving outer disk and a stellar component. The possible presence and properties of a stellar cluster will be addressed in Paper~II.

\section{Conclusion}
\label{sec:conclusion}

We fit the soft-state portions of the HLX--1 outbursts with a relativistic thin-disk time-dependent model. The \textit{Swift}/XRT light curves are fit simultaneously across epochs, with the black-hole and disk-geometry parameters shared between outbursts and the injected mass and timing parameters allowed to vary from epoch to epoch. We also include the \textit{HST} UV/optical/IR data contemporaneous with the soft X-ray state in Epoch~3 as a broadband constraint on the disk spectrum. Our main conclusions are as follows.

\begin{itemize}

\item The fitted soft-state portions of the outbursts are consistent with a common global solution. We infer an intermediate-mass black hole, $\log_{10}(M_{\bullet}/M_\odot)=4.5\pm0.2$, and a large characteristic disk-formation scale, $\log_{10}(r_0/r_g)=3.8\pm0.3$. The free-$r_0$ analysis shows no systematic epoch-to-epoch evolution in this scale and leaves the remaining shared parameters unchanged.

\item The contemporaneous Epoch~3 emission, from the UV/optical/IR bands to X-rays, is well described by the disk-only model during the soft state.
The broader multi-epoch UV/optical/IR emission and possible stellar-cluster component will be studied in Paper~II.

\item The outburst-to-outburst diversity is captured mainly by changes in the epoch-dependent parameters $(M_{{\rm d},i}, t_{{\rm visc},i}, t_{0,i})$. Freeing $r_0$ per epoch shifts these constraints only modestly ($<1\sigma$).
This separation shows that the fitted soft-state diversity can be reproduced with a single accretion geometry and different injected masses and viscous responses.

\item The inferred injected masses cluster around $M_{\rm d}\sim10^{-3}\,M_\odot$ per outburst. This fuel requirement can be supplied by repeated partial stripping of a low-mass donor, while the shared circularization scale maps to a highly eccentric, grazing encounter geometry.

\end{itemize}

Together, the repeated injected-mass scale, the shared angular-momentum scale, and the fuel-budget estimate support a repeating partial tidal disruption interpretation for the fitted soft-state evolution of HLX--1.

\begin{acknowledgments}
We thank the anonymous referee for comments that improved the clarity of the manuscript.
F.Y. thanks Eliot Quataert for insightful discussions.
A.M. acknowledges support from the Ambrose Monell Foundation, the W.M. Keck Foundation and the John N. Bahcall Fellowship Fund at the Institute for Advanced Study.
The calculations presented in this article were performed on computational resources managed and supported by Princeton Research Computing, a consortium of groups including the Princeton Institute for Computational Science and Engineering (PICSciE) and the Office of Information Technology’s High Performance Computing Center and Visualization Laboratory at Princeton University.
Some of the data presented in this article were obtained from the Mikulski
Archive for Space Telescopes (MAST) at the Space Telescope Science Institute.
The specific observations analyzed can be accessed via
\dataset[doi:10.17909/bbxa-vf88]{https://doi.org/10.17909/bbxa-vf88}.
\end{acknowledgments}

\facilities{HST, Swift.}

\software{
\texttt{FitTeD} \citep{fitted},
\texttt{emcee} \citep{ForemanMackey+2013}.
          }

\bibliography{HLX-1-1}{}
\bibliographystyle{aasjournalv7}

\end{document}